\documentclass{iau}
\usepackage{graphicx}

\title[Characterisation of M giant variability with {\em Kepler}] 
{Variability of M giant stars based on {\em Kepler} photometry: general  characteristics}

\author[B\'anyai et al.]{E. B\'anyai; L. L. Kiss \& members of KASC WG12}

\affiliation{Konkoly Observatory, Hungarian Academy of Sciences, \\ H-1121 Budapest, Konkoly Thege M. \'ut 15-17, Hungary \\ email: {\tt ebanyai@konkoly.hu}
}

\pubyear{2013}
\volume{301}  
\pagerange{119--126}
\jname{Precision Asteroseismology}
\editors{A.C. Editor, B.D. Editor \& C.E. Editor, eds.}
\begin{document}

\maketitle

\begin{abstract}
Our study based on the continuous, high-precision observations covering more than three years provided by the Kepler mission reveals a new systematic  effect in the data,  a possible transition region between solar-like and Mira-like oscillations, and gives an overview of M giant variability on a wide range of time-scales (hours to years). 
\end{abstract}

\firstsection 
\section{Introduction}
 
M giants are among the longest-period pulsating stars, hence their studies have traditionally been restricted to analyses of low-precision visual observations, or more recently, accurate ground-based CCD data. 

We used thirteen quarters of Kepler long-cadence observations (one point per every 29.4 minutes) to analyse M giant variability, with a total time-span of over 1000 days. 
About two-thirds of our 317 stars have been selected from the ASAS-North survey of the Kepler field, with the rest supplemented from a randomly chosen M giant sample. 

\section{A new systematic effect -- {\em Kepler}-year in the data}

We found that the 23\% of the total sample has small variations with sinusoidal modulation and period similar to the Kepler-year (372.5 days). Periods and phases indicated that those changes are likely to be caused by a so-far unrecognised systematics in the Kepler data.
The phase of the Kepler-year variations varies among these stars. Further investigation revealed that the positions of the stars in the CCD-array display a clear correlation with the phase.

\section{General Characteristics}

Based on the complexity in the time and the frequency domains, we sorted the stars into three groups. Stars in Group 1 have a wide range of periods between a few days and 100 days. Group 2 contains stars with very low-amplitude light curves that are mostly characterised by short-period oscillations, occasionally supplemented by slow changes that may be related to rotational modulation or instrumental drifts. Stars with light curves containing only a few periodic components (Miras and SRs) compose Group 3.

\begin{figure}
\begin{center}
\includegraphics[width=3.4in]{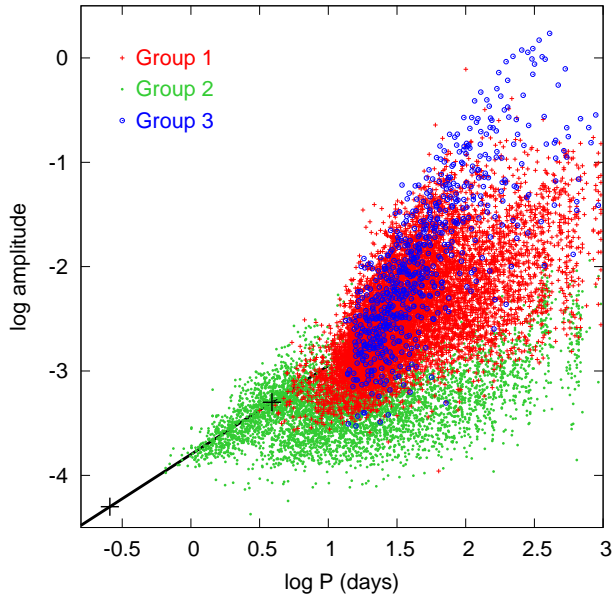} 
 \caption{The period-amplitude relations for the whole sample using all the 50 frequencies and amplitudes. Different symbols distinguishes the three groups. Plus signs refer to two selected points in the top panel of Fig. 3. of \cite{Huber2011} related to solar-like oscillations. The black line is drawn through these points.}
\label{image:PA-relation}
\end{center}
\end{figure}

The power spectra and wavelets reveal very complex structures and rich behaviour. Peaks in the spectra are often transient in terms of time-dependent amplitudes. The overall picture is that of random variations presumably related to the stochasticity of the large convective envelope.

Determination of the significant period ratios provided the following results. The most populated clump is around $P_{\rm short}/P_{\rm long} \approx $0.7-0.8, a ratio  that is known to belong to the upper RGB stars. 
There is another distinct clump around period ratios of 0.5 that could be related to pulsation in the fundamental mode and first radial overtones \cite{Takayama2013}. The vertical concentrations at log $P \approx$ 2.57 and 2.87 result from the Kepler-year variability

In Fig. \ref{image:PA-relation} the bulk of the giants are spread in a triangular region.
To the left of this upper envelope there is a distinct feature which shows strong correlation between the period and amplitude. To validate that the correlation is indeed in the extension of the $\nu_{\rm max}$-amplitude relation for the solar-like oscillations, we added two points, marked by the large plus signs, and a line drawn through these points. These points are taken from the top panel of Fig. 3 of \cite{Huber2011}, where the oscillation amplitude vs. $\nu_{\rm max}$ is shown for their entire Kepler sample.
The excellent agreement between the line and the period-amplitude relation for Group 2, indicates that these stars are indeed the long-period extension of the solar-like oscillations.
 
\section*{Acknowledgements}
This project has been supported by the Hungarian OTKA Grants K76816, K83790, K104607 and HUMAN MB08C 81013 grant of Mag Zrt., and the Lend\"ulet-2009 Young Researchers Program of the Hungarian Academy of Sciences. The research leading to these results has received funding from the European Community's Seventh Framework Programme (FP7/2007-2013) under grant agreement no. 269194

\end{document}